\documentclass[aps,prd,superscriptaddress,nofootinbib,letter]{revtex4}
\usepackage[latin1]{inputenc}
\pdfoutput=1
\usepackage{ifpdf}
\usepackage{pgf}
\usepackage{graphicx}
\usepackage{epstopdf}
 \usepackage{latexsym}
\begin{document}
\title{$\gamma$ gravity: Steepness control}
\author{M\'arcio O'Dwyer}
\email{modwyer@if.ufrj.br}
\author{S\'ergio E. Jor\'as}
\email{joras@if.ufrj.br}
\author{Ioav Waga}
\email{ioav@if.ufrj.br}
\date{\today}
\affiliation{Instituto de F\'\i sica, Universidade
Federal do Rio de Janeiro\\{C. P. 68528, CEP 21941-972, Rio de Janeiro, RJ,
Brazil}}


\begin{abstract}

We investigate a simple generalization of the metric exponential $f(R)$ gravity theory that is cosmologically viable and compatible with solar system tests of gravity. We show that, as compared to other viable $f(R)$ theories, its steep dependence on the Ricci scalar $R$ facilitates agreement with structure constraints, opening the possibility of $f(R)$ models with equation-of-state parameter that could be differentiated from a cosmological constant ($w_{de}=-1$) with future surveys at both background and perturbative levels.

\end{abstract}

\maketitle

\section{Introduction}

One of the major puzzles in modern cosmology is to unveil the physical mechanism responsible for the late-time cosmic acceleration. The two main approaches considered in the literature are the following: (1) the existence of an unknown component with negative pressure, generically denominated dark energy,  whose equation-of-state parameter ($w=p/\rho$) satisfies $w<-1/3$, and (2) general relativity (GR) has to be modified at large scales or, more accurately, at low curvatures (modified gravity). The simplest dark-energy candidate is the cosmological constant ($\Lambda$) that, as it is well known, faces some theoretical difficulties (such as its tiny value when comparing the theoretical expectation to the vacuum energy density, the so-called cosmic coincidence and its fine-tuning), but it is in very good accordance with current cosmological observations. The simplest modified-gravity candidate is the so-called $f(R)$ gravity in which the Lagrangian density ${\cal L}=R+f(R)$ is a nonlinear function of the Ricci scalar $R$. 

An accelerated expansion appears naturally in $f(R)$ theories. Indeed, the very first inflationary model, proposed more than 30 years ago \cite{Staro80}, is curvature driven by a term proportional to the square of the Ricci scalar $R$ [$f(R)=\alpha R^2$ ($\alpha>0$)] and, interestingly enough, it is still in excellent accordance with current observations \cite{Ade}. More recently the same idea has been explored in Refs. \cite{Capoz} and \cite{Carrol}, but in the low-curvature regime. They considered a late-time acceleration driven by an inverse power law, $f(R)=-\alpha R^{-n}$ ($\alpha>0$ and $n>0$). However, those models have a serious drawback: they do not present a regular matter-dominated phase.  The scale factor of the Friedman-Robertson-Walker (FRW) metric, $a$, grows with cosmic time, $t$, as $a \propto t^{1/2}$ (instead of the standard $a \propto t^{2/3}$) and, therefore, those models are incompatible with structure formation \cite{Amendola}.

There are viable $f(R)$ gravity theories that do not present the above mentioned problem and satisfy both local gravity and cosmological constraints \cite{Hu, Staro, Appleby,Cognola,Linder}. These models suffer, however, another difficulty: the curvature singularity in cosmic evolution at a finite redshift \cite{Frolov}. Indeed, this seems to be a generic feature of all the so-called ``viable'' $f(R)$ theories \cite{Thongkool}, including the new one proposed in the present paper. However, this singularity problem can be cured, for instance, by adding to the density Lagrangian a high curvature term proportional to $R^2$ \cite{Appleby10}. Therefore, it appears that it is not possible to have cosmic acceleration with a totally consistent $f(R)$ theory modifying gravity {\it only} at low curvatures. In this work we are not addressing this issue and we will be concerned only with $f(R)$ modifications at low curvatures.

Another feature of all viable $f(R)$ theories discussed so far in the literature is that structure formation imposes such strong constraints on their parameters that their cosmic expansion history, in practice, cannot be discriminated from that of cold dark matter with a cosmological constant ($\Lambda$CDM). In this work we propose a class of $f(R)$ theories that may change this characteristic. The proposed modification depends on a parameter that controls the steepness of $f(R)$ allowing  measurable (in the near future) deviations  from $\Lambda$CDM at both perturbation and background levels ($|\Delta w| \sim 2 - 4 \%$), while still compatible with both current observations.

This paper is organized as follows: in Sec. \ref{model}, we introduce the model, obtain its field equations, investigate its background expansion history and examine the behavior of the effective equation of state parameter as a function of the parameters of the model and redshift. In Sec. \ref{local}, we discuss the constraints on the model from local gravity tests and growth of structure. Our conclusions are presented in Sec. \ref{conclusions}.

\section{The $\gamma$-Gravity Model: Field Equations and Expansion History} \label{model}

\begin{figure}[t]
    \includegraphics[width=9.0cm,height=5.5cm]{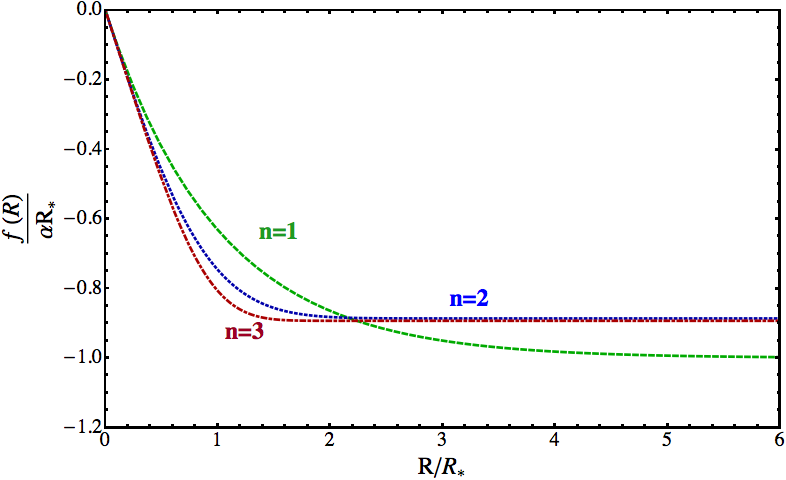}
    \caption{Behavior of $f(R)$ for $n=1,2$ and $3$. The higher the $n$, the faster the function reaches its high-curvature limit. Accordingly, one recovers GR with an effective cosmological constant.}
    \label{fig:1}
\end{figure}

In this work we investigate gravity theories described by the following action
\begin{equation}
    S=\int {d^{4}x\sqrt{-g}}\left[ \frac{1}{16\pi G}\left( R+f (R)\right)+ \mathcal{L}_{mat}
    \right] ,  \label{acao}
    \vspace{-0.1cm}
\end{equation}
where $f(R)$
is an arbitrary function of the Ricci scalar  $R$. General relativity with a
cosmological constant is obtained in the special case in which 
$f (R)=-2\Lambda =$ const.
Here we are interested in $f(R)$ gravity theories described by the following ansatz
\begin{equation}
 f(R)=  - \frac{\alpha R_{\ast}}{n} \gamma\left(\frac1n, (\frac{R}{R_{\ast}})^n\right)  ,\label{fR}
\end{equation}
where $\gamma(a,x):=\int_0^x e^{-t} t^{a-1}dt$ is the lower incomplete gamma function \cite{wolfram} and $\alpha$, $n$ and $R_{\ast}$ are free positive parameters. We call this theory generalized exponential gravity or $\gamma$ gravity for short. It is straightforward to verify that (\ref{fR}) generalizes several interesting cases. For instance, if $n=1$ we obtain exponential gravity \cite{Cognola}\cite{Linder}
\begin{equation}
f(R)=-\alpha R_{\ast} (1-e^{-R/R_{\ast}}),
\end{equation}
while for $n=2$ we get
\begin{equation}
      f(R)=-\alpha R_{\ast} \frac{\sqrt{\pi}}{2} \text{Erf}(\frac{R}{R_{\ast}}), 
      \end{equation} 
where $\text{Erf}(x)$ is the error function. It follows from (\ref{fR}), that the first derivative of  $f(R)$ with respect to $R$ is given by,
\begin{equation}
f_R:= \frac{d f}{{d}R}= -\alpha e^{-(R/R_{\ast})^n} \label{DR}
\end{equation}
and the second derivative by
\begin{equation}
f_{RR}:= \frac{d^{2}f}{{d}
R^{2}}= \frac{n \alpha}{R_{\ast}} e^{-(R/R_{\ast})^n}(\frac{R}{R_{\ast}})^{n-1}. \label{DRR}
\end{equation} 
For high curvatures ($R \gg R_{\ast}$), it follows from (\ref{fR}) that GR with $\Lambda$ is recovered, although there is no cosmological constant [$f(0)=0$]. As we will discuss further down, the fact that in $\gamma$ gravity $f_R$ is proportional to $e^{-(R/R_{\ast})^n}$ is crucial to facilitate agreement of the proposed $f(R)$ with structure formation constraints. Clearly as we increase $n$, the steepness of $f(R)$ increases (Fig.\ref{fig:1}). Therefore, if $n>1$ the steepness is larger than in the exponential case which implies that, cosmologically, when we go back in time from the present (increasing $R$), the $\Lambda$CDM regime is achieved faster. 
      
\begin{figure}[t]
    \includegraphics[width=9.0cm,height=5.5cm]{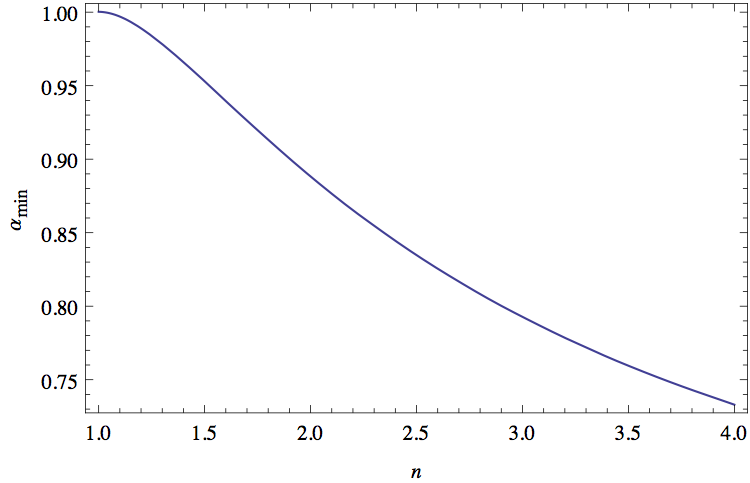}
    \caption{Minimum value of $\alpha$ as a function of $n$ required to have a final de Sitter attractor. For the figure $\tilde{\Omega}_{m0}=0.28$ is assumed.}
    \label{fig:2}
\end{figure}

      In principle, the ansatz (\ref{fR}) can satisfy all the $f(R)$ stability conditions  \cite{Pogosian}: (a) $f_{RR} >0$ (no tachyons); (b) $1+f_{R}>0$ [the effective gravitational constant ($G_{eff} = G/(1+f_R)$] does not change sign (no ghosts)); 
(c) $\lim_{R\rightarrow \infty }f /R=0$ and
$\lim_{R\rightarrow \infty }f _{R}=0$ (GR is recovered at early
times); and d) $|f_R|$ is small at recent epochs (to satisfy solar and galactic scale constraints). Furthermore, Eq.~(\ref{fR}) can also satisfy cosmological viability criteria \cite{Amendola07}.  We characterize a viable cosmological model as one that starts at a radiation-dominated phase and has a saddle-point matter-dominated phase followed by an accelerated expansion as a final attractor. Formally, such criteria can be stated by using the parameters  $\bar{m}:=Rf_{,RR}/(1+f_{R})\;$ and $\;\mathfrak{r}
:=-R(1+f_{R})/(R+f)\,$. An early matter-dominated epoch of the Universe can be achieved if $ \bar{m}(\mathfrak{r} \approx -1) \approx 0^+$ and $\bar{m}/\mathfrak{r}(\mathfrak{r}\approx -1) > -1$. Furthermore, a necessary condition for a late-time accelerated attractor is $0<\bar{m}(\mathfrak{r}\approx -2)\leq 1$. As a consequence, for fixed $n$, there is a minimum value ($\alpha_{min}$) of the parameter $\alpha$ such that for values $\alpha > \alpha_{min}$ the latter constraint can be satisfied. Figure \ref{fig:2} displays $\alpha_{min}$ as a function of $n$ for a chosen fixed value of $\tilde{\Omega}_{m0}$ (see definition bellow).

      By varying the action (\ref{acao}) with respect to the metric we obtain the modified Einstein equations:
\begin{equation}
   \left(1+ f_{R}\right)R_{\mu \nu }-\frac12 g_{\mu \nu }\left(R +f\right) +\left( g_{\mu \nu }\Box - \nabla _{\mu }\nabla _{\nu}\right) f_{R}=8\pi GT_{\mu \nu }.  \label{eqf}
\end{equation}
For a homogeneous universe filled with matter energy density $\bar{\rho}_m$ and radiation energy density $\bar{\rho}_r$ we use the above equation to get the modified Friedman equation,
      \begin{equation}
    H^{2} +\frac{f}{6} -f_R \left(H^2+H H^\prime\right) +H^2 f_{RR} R^\prime =  \frac{8\pi G}{3} \bar{\rho}, \label{H2}
\end{equation}
where $^\prime :=d/dy$ ($y=\ln a$), $H:=\dot{a}/a$ is the Hubble parameter (a dot denotes the derivative with respect to cosmic time), $\bar{\rho}=\left(\bar{\rho} _{m}+\bar{\rho} _{r}\right)$ and we assume flat space. For the FRW background we have
\begin{equation}
R=12 H^2 + 6 H H^\prime. \label{R}
\end{equation} 
To solve the above equations, we follow \cite{Hu} (see also \cite{Bamba} and \cite{Linder}) and introduce the following variables:
\begin{equation}
x_1(y)=\frac{H^2}{m^2}-e^{-3y} - d - a_{eq} \;e^{-4y}, \label{x1}
\end{equation}
\begin{equation}
x_2(y)=\frac{R}{m^2}-3 e^{-3y} - 12\left(d +x_1(y)\right), \label{x2}
\end{equation}
where $d:=\frac{\alpha R_{\ast} \Gamma(1/n)}{6 n m^2}$, $a_{eq}=\bar{\rho}_{r0}/\bar{\rho}_{m0} \simeq 2.9 \times 10^{-4}$,  $m^2:=\frac{8\pi G}{3} \bar{\rho}_0$ and, as usual, a quantity with a subscript ``0'' denotes its value at present time. Above, $\Gamma (x)$ is the gamma function. Since $\bar{\rho}_{r0} \ll \bar{\rho}_{m0}$, from now on we assume $m^2 = \Omega_{m0} H_0^2$. With the above definitions we obtain
\begin{equation}
x^\prime_1(y)=\frac{x_2(y)}{3}, \label{x1p}
\end{equation}
\begin{equation}
x^\prime_2(y)=\frac{R^\prime}{m^2} + 9 e^{-3y} - 4 x_2 (y), \label{x2p}
\end{equation}
where
\begin{equation}
\frac{R^\prime}{m^2}=\frac{e^{-3y}+a_{eq} \;e^{-4y}}{H^2 f_{RR}} -\frac{1}{m^2 f_{RR}}\left(1+\frac{f}{6 H^2}\right)+\frac{f_R}{m^2 f_{RR}}\left(\frac{R}{6 H^2}-1\right). \label{Rprime}
\end{equation}
Again, the definitions (\ref{x1}) and (\ref{x2}) should be used above to eliminate $H^2$ and the dependence on $R$ in Eq.~(\ref{Rprime}). 

\begin{figure}[t]
\begin{centering}
    \includegraphics[width=0.49\textwidth]{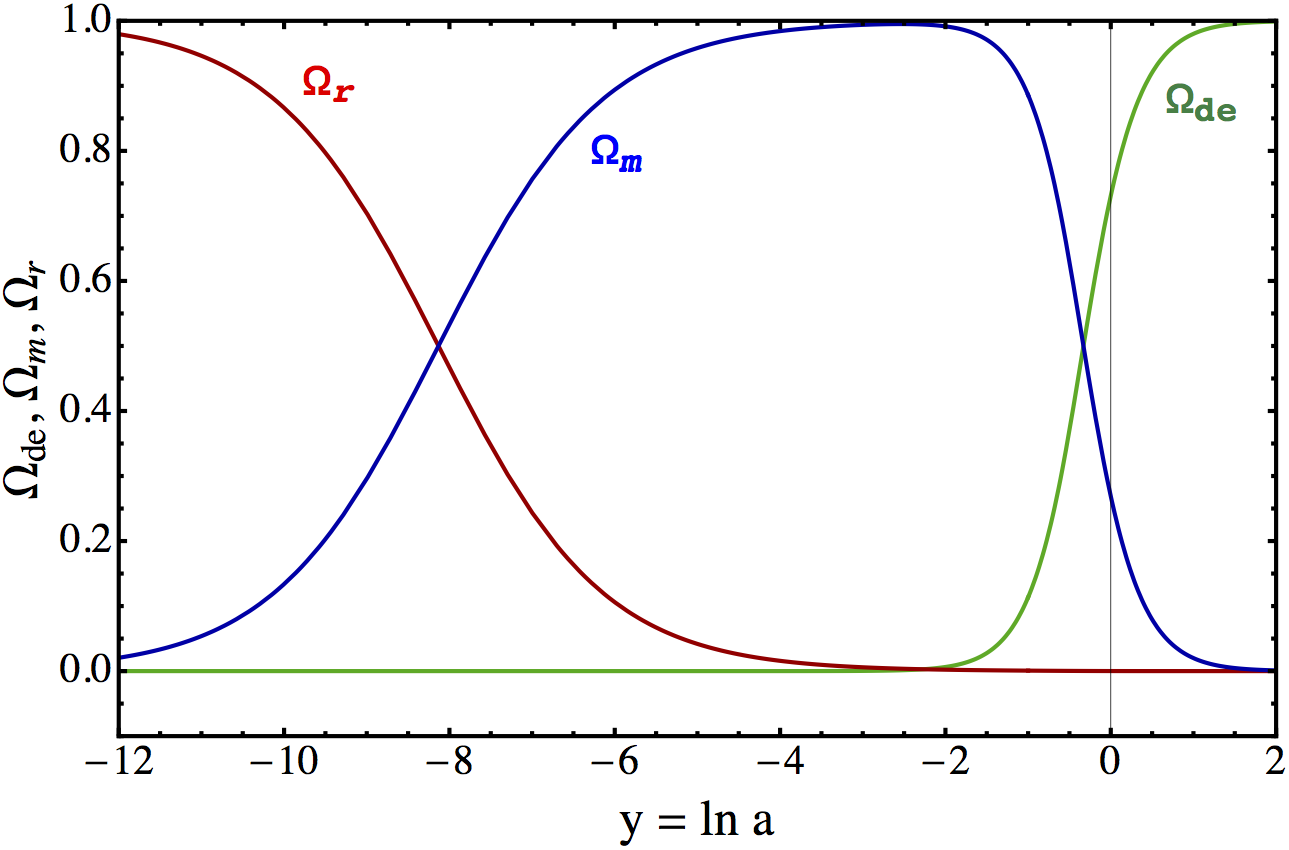}
     \includegraphics[width=0.49\textwidth]{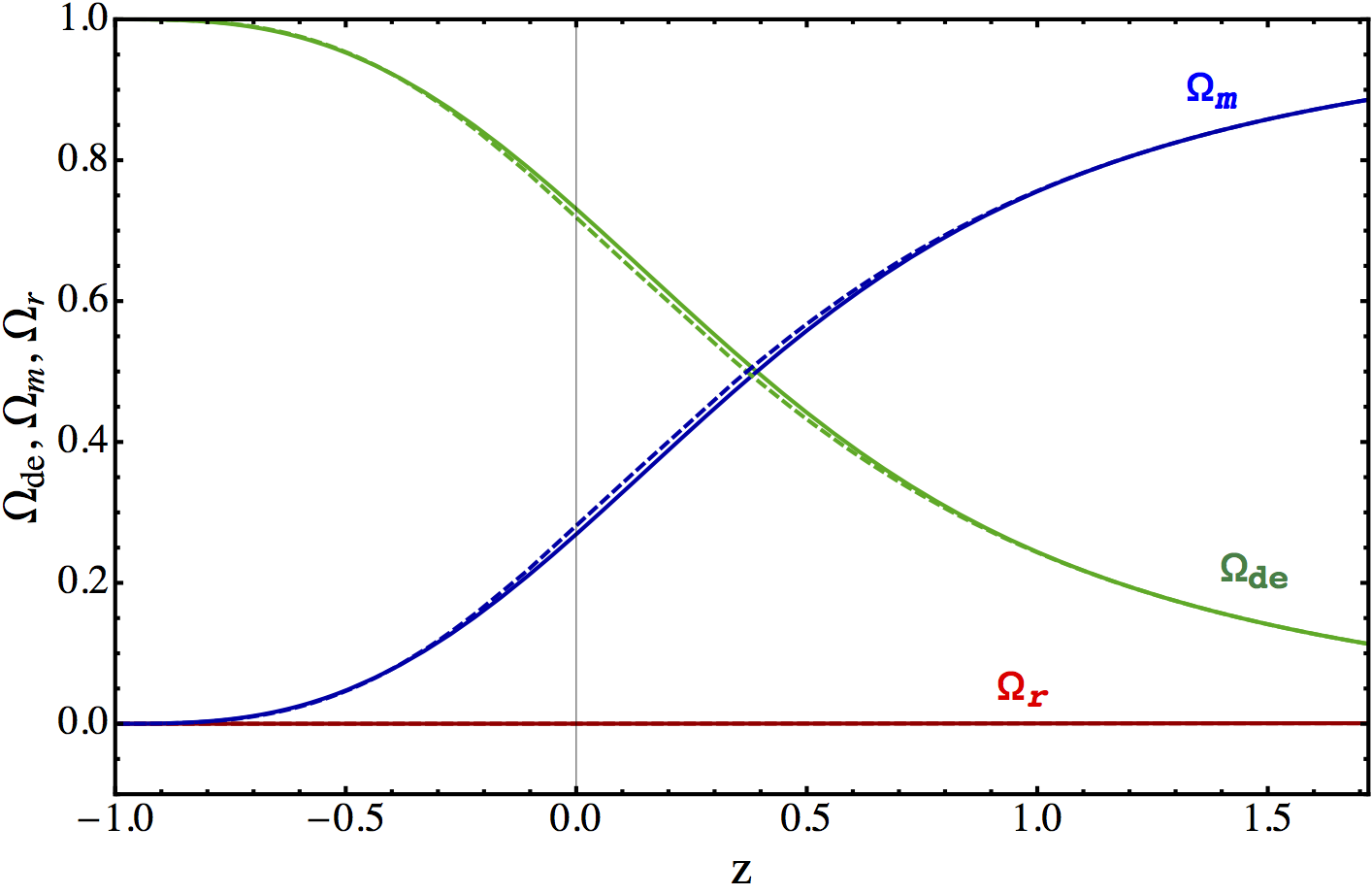}
      \par \end{centering}
    \caption{Fractional energy densities $\Omega_{de}$ (green curves), $\Omega_{m}$ (blue curves) and $\Omega_{r}$ (red curves) as a function of  $y:=\ln a$ (left panel) and $z$ (right panel) with $n=2$, $\alpha=1$ and $\tilde\Omega_{m0}=0.28$. Note the dashed lines, visible only in the right panel, indicating the corresponding quantities for the reference $\Lambda$CDM model.}
    \label{fig:3}
\end{figure}

\begin{figure}[t]
    \includegraphics[width=0.49\textwidth]{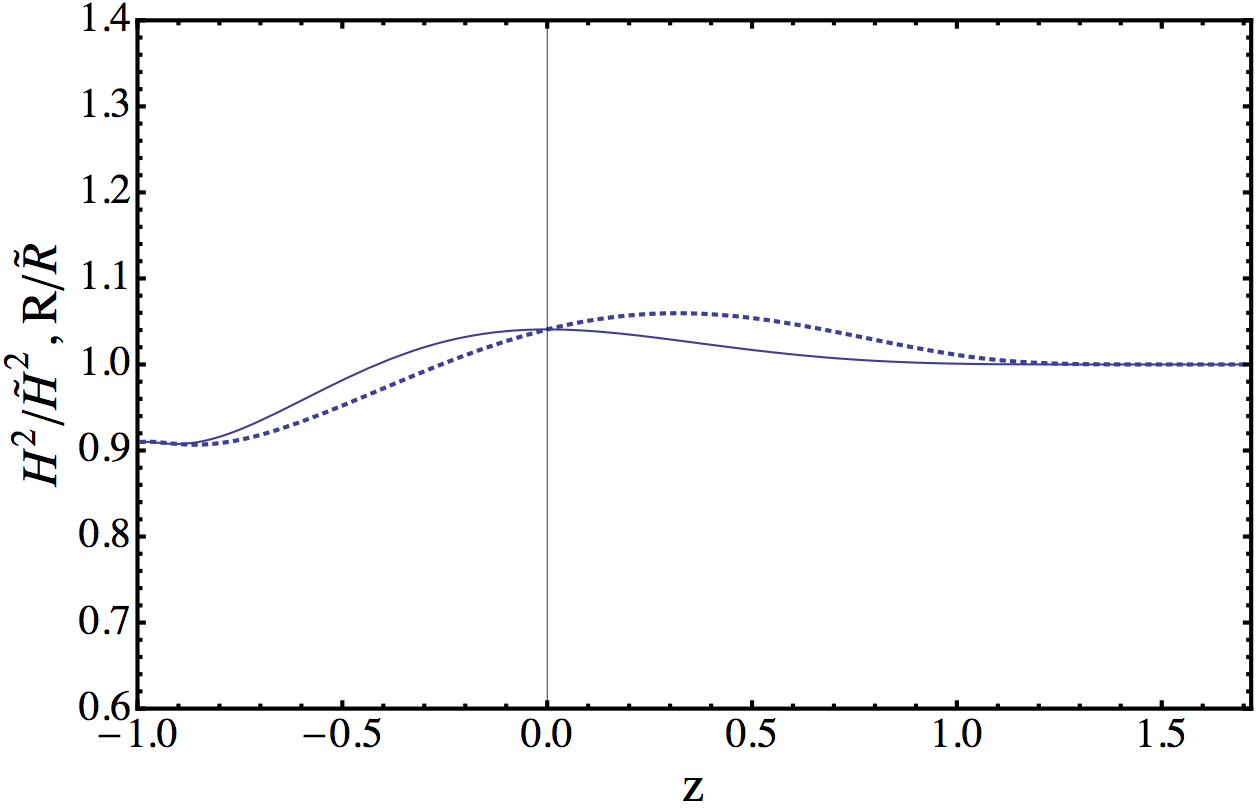}
     \includegraphics[width=0.49\textwidth]{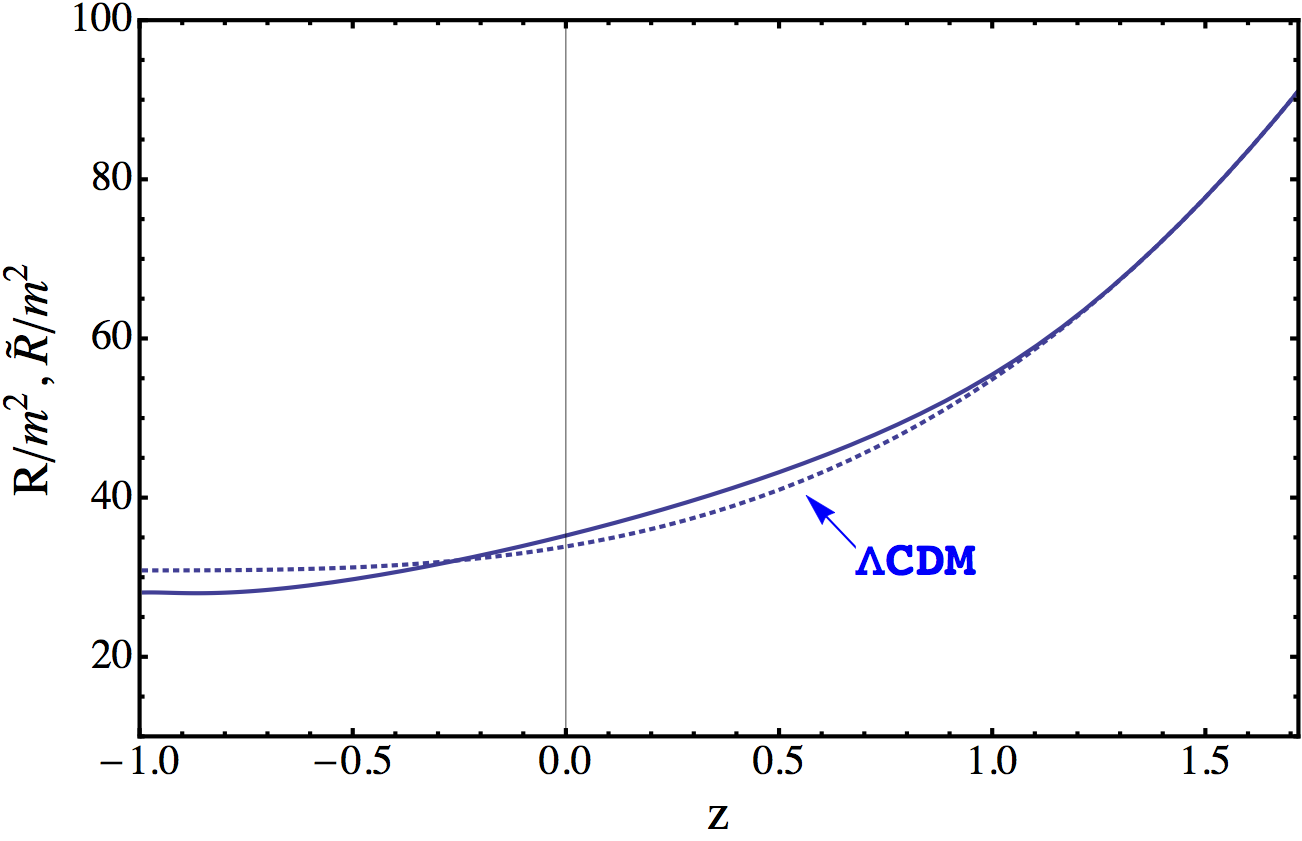}
    \caption{Comparison among quantities in $\Lambda$CDM (symbols with tildes) and in the present model with $n=2$, $\alpha=1$ and $\tilde\Omega_{m0}=0.28$.  Left panel: ratios $H^2/\tilde H^2$ (solid) and $R/\tilde R$ (dotted line) as a function of $z$. Right panel: comparison of $R/m^2$ and $\tilde{R}/m^2$ (dotted line).}
    \label{fig:4}
\end{figure}

Each model is characterized by fixed values of the parameters $\alpha$, $n$ and $R_{\ast}$. Since at high curvature, when $R \gg R_{\ast}$, the models behave like $\Lambda$CDM, by using that $\lim_{x\rightarrow \infty}\gamma\left(\frac1n, x \right)=\Gamma (1/n)$,  from Eq.(\ref{fR}), at this limit, we get $2\tilde{\Lambda}=\alpha R_{\ast} \Gamma(1/n)/n$) and we can write $R_{\ast}$ as 
\begin{equation}
 R_{\ast}=\frac{6 n m^2}{\alpha\Gamma(1/n)}\frac{1-\tilde{\Omega}_{m0}}{\tilde{\Omega}_{m0}}. \label{Rstar}
 \end{equation}
In this work, in our numerical computation, we always assume that $\tilde{\Omega}_{m0}=0.28$. Here $\tilde{\Omega}_{m0}$ represents the present value of the matter density parameter that a $\Lambda$CDM model would have, if it had the same matter density $\bar{\rho}_{m0}$ as the modified gravity $f(R)$ model. As a consequence, if $\tilde{H}_0$ is the Hubble constant in the reference $\Lambda$CDM model, we should have $ \tilde{\Omega}_{m0}\tilde{H}_0^2=\Omega_{m0}H_0^2$. From (\ref{Rstar}) we also get that $d =(1-\tilde{\Omega}_{m0})/\tilde{\Omega}_{m0}\simeq 2.57$ and $ R_{\ast}/m^2 = (15.43, 17.41, 17.28)$ for $\alpha=1$ and $n=1,2,3$, respectively.
To solve the system given by Eqs. (\ref{x1p}) and (\ref{x2p}) we use the initial condition that, at high curvature, $x_1(y_i) = x_2(y_i)=0$, where $y_i<0$, is an initial value of $y=\ln a$. For $y<y_i$ the solution is matched to the reference $\Lambda$CDM model (one with the same $\bar{\rho}_{m0}$). It is straightforward to verify that, as defined,  $x_1$ and $x_2$ are always zero during the $\Lambda$CDM phase. From $x_1(y)$ and $x_2(y)$, several quantities can be obtained. For instance, the effective dark energy equation of state ($w_{de}$) is given by,
\begin{equation}
w_{de} = -1 - \frac{1}{ 9} \frac{x_2}{x_1 + d}.
\end{equation}
We also have
\begin{equation}
\Omega_{de}(y) =\frac{x_1+ d}{d + x_1 + e^{-3 y} + a_{eq} e^{-4y}},
\end{equation}
\begin{equation}
\Omega_{r}(y) =\frac{a_{eq} e^{-4y}}{d + x_1 + e^{-3 y} + a_{eq} e^{-4y}}
\end{equation}
and $\Omega_m=1-\Omega_{de}-\Omega_{r}$. Figure \ref{fig:3} (left panel) shows the evolutions of $\Omega_{r}$, $\Omega_{m}$ and $\Omega_{de}$  as functions of $y$ for the case $n=2$, $\alpha=1$ and $\tilde{\Omega}_{m0}=0.28$. For the same model, the evolution of the same quantities, as functions of redshift, are displayed in the right panel of the figure, together with the corresponding quantities in the reference $\Lambda$CDM model (dashed curves). Note that $\Omega_{m0}$ is slightly smaller than $\tilde{\Omega}_{m0}$. This occurs because from  $z \sim 1.5$ (when deviation from $\Lambda$CDM starts to become relevant for the cosmic expansion) on until $z\sim 0$, $H(z)$ is always slightly larger than $\tilde{H}$ and $\Omega_m \propto H^{-2}$.  The ratios $H^2/\tilde{H}^2$ and $R/\tilde{R}$ as a function of redshift are displayed in the left panel of Fig. \ref{fig:4} for the same models as before. The right panel of the same figure shows $R/m^2$ and $\tilde{R}/m^2$ as a function of $z$.

\begin{figure}[t]
    \includegraphics[width=9.0cm,height=5.5cm]{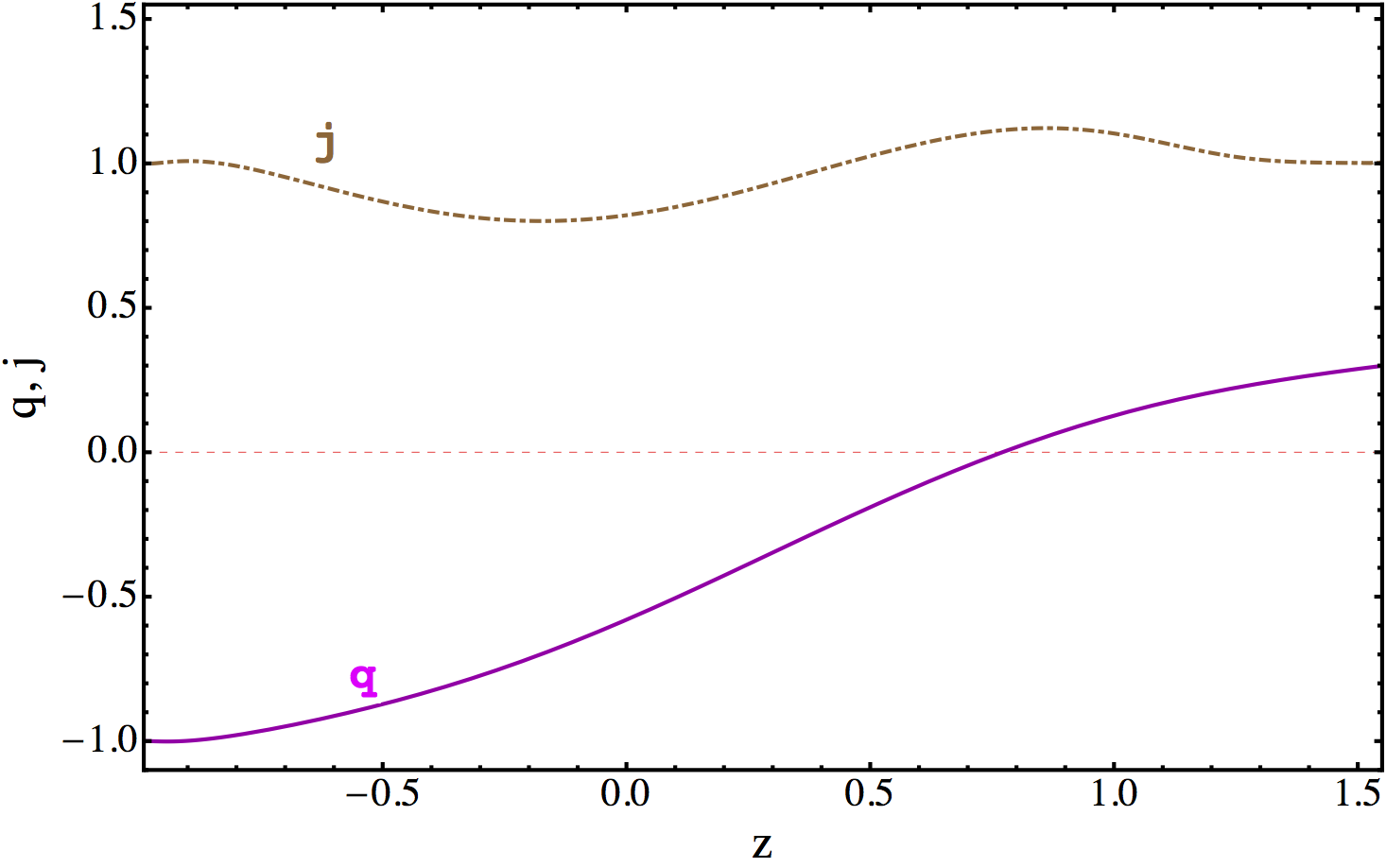}
    \caption{Evolution of the deceleration parameter $q$ and the jerk $j$ with the redshift $z$ with the same parameters used in the previous plots. Deviations from $\Lambda$CDM start at $z \sim 1.5$ when $j$ deviates from unity.}
    \label{fig:5}
\end{figure}

Figure \ref{fig:5} shows the evolution of the deceleration parameter ($q\equiv -\frac{a\ddot{a}}{\dot{a}^{2}}$) and the jerk ($j\equiv \frac{a^{2}\dddot{a}}{\dot{a}^{3}}$) as functions of $z$, in the same case as above. We can compare the values of $z_t$, the transition redshift [$q(z_{t}]=0$) from decelerated to accelerated expansion, and the parameter $\tau$, related to the width of the transition [$\tau^{-1}=\frac32 j(z_t)$] \cite{Giostri}, in the considered $f(R)$ case and the corresponding fiducial  $\Lambda$CDM model.  For the latter we obtain $\tau=1/3$ and $z_t=0.73$ while for the former we get $\tau=0.30$ and $z_t=0.77$. This happens because when $f(R)$ starts to become effective--- that is, when $R$ starts to become comparable to $R_{\ast}$ ---the Universe enters in a phantom phase ($w_{de}<-1$) that accelerates the transition and reduces its width.

\begin{figure}[t]
\begin{centering}
    \includegraphics[width=0.49\textwidth]{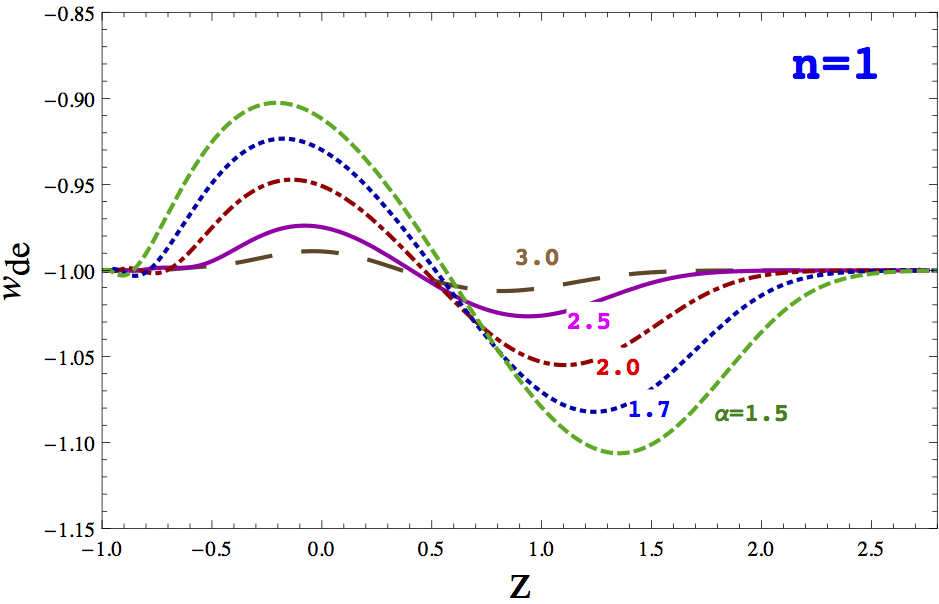}
        \includegraphics[width=0.49\textwidth]{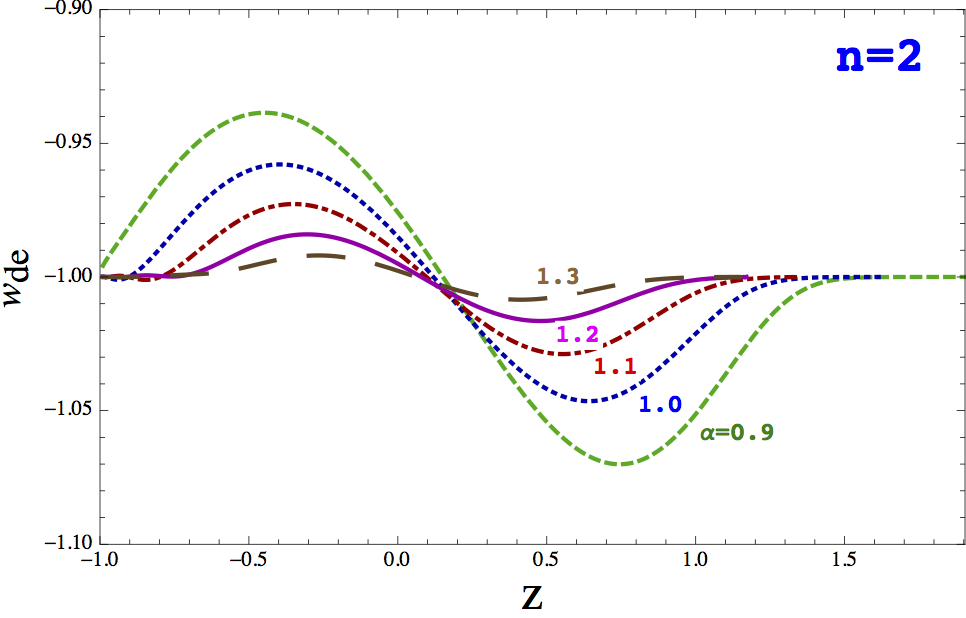}
            \includegraphics[width=0.49\textwidth]{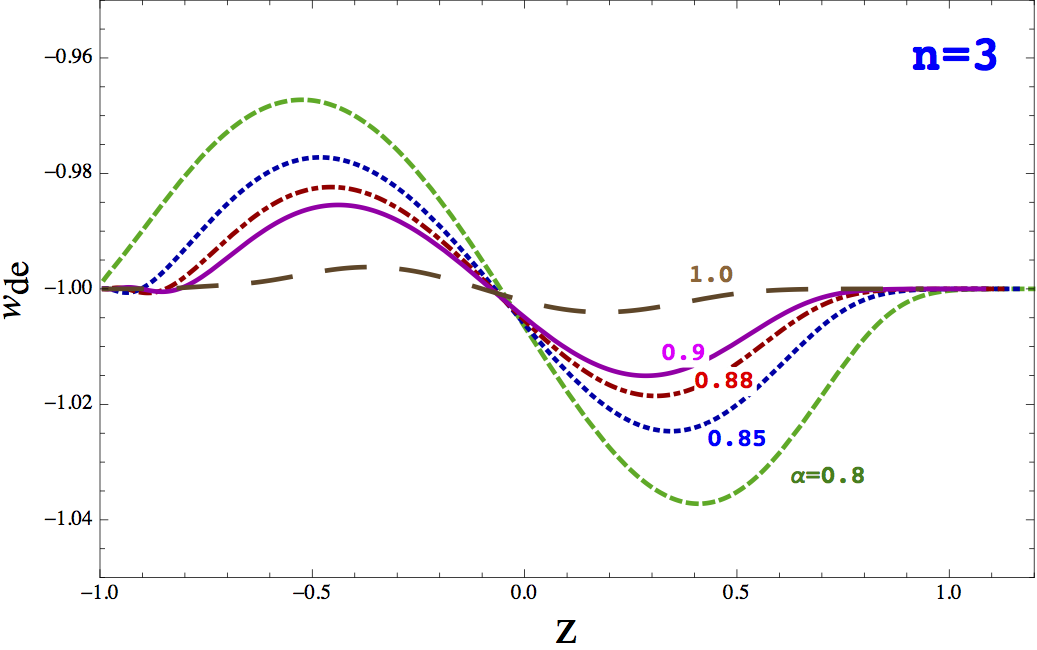}
           \par \end{centering}
    \caption{Effective equation-of-state parameter $w_{de}$ as a function of $z$ for $n=1,2$ and $3$ and different values of $\alpha$. Note the different ranges in both axes of the three panels. The higher the $n$, the smaller the redshift where $w_{de}$ deviates from $-1$ and the smaller the amplitude of the deviation itself for the same $\alpha$. For fixed $n$, increasing $\alpha$ also decreases the amplitude.}
    \label{fig:6}
\end{figure}

Figure \ref{fig:6} displays the effective dark-energy equation of state as a function of the redshift for models with $n=1$, $2$ and $3$ and different values of the parameter $\alpha$. For high redshift values (when $R \gg R_{\ast}$), $w_{de} = -1$ and, as expected, all the models behave like $\Lambda$CDM at early times. For $z \rightarrow -1$ we also have $w_{de} \rightarrow -1$ indicating that, asymptotically ($t \rightarrow \infty$), the models have a de Sitter final attractor. Note that, for fixed $n$, as the parameter $\alpha$ increases the models approach the $\Lambda$CDM expansion history behavior, \textit{i.e.}, $|1+w_{de}|_{max}$ decreases. For instance, if $n=1$ and $\alpha =3$, we get $|1+w_{de}|_{max}\sim 10^{-2}$, while for $\alpha=6$, we obtain $|1+w_{de}|_{max} \lesssim 10^{-4} $. The redshift of phantom crossing also decreases with increasing $\alpha$ for fixed $n$. Note also that models with larger values of  $n$ (higher steepness) start the phantom phase later in the Universe evolution. For instance, if $n=1$ this deviation occurs at $z \lesssim 2.5$, while for $n=2$ it occurs at $z \lesssim 1.5$,  and for $n=3$ at $z \lesssim 1.0$.  As we will discuss in the next section, having the transition from $w_{de}=-1$ to the phantom phase at  $z \lesssim 1.0$ is interesting because it can ease the agreement with structure formation constraints. Note from Fig. \ref{fig:6} that maximum deviations in the equation-of-state parameter for models with $n=3$  occur at $z \sim 0.5$, the redshift at which future surveys like WFIRST have their best sensitivity \cite{Spergel}.  Since the maximum value of $|1+w_{de}|_{max}$  also decreases with increasing $n$, if we are looking for models with relative high values ($2\%-4\%$) of $|1+w_{de}|_{max}$ at $z\sim 0.5$, the steepness parameter cannot be much larger than $n=3$.

\section{Local Tests, Structure Formation and Power Spectrum} \label{local}

Constraints on $f(R)$ gravity theories from solar-system tests and equivalence principle violation have been discussed by several authors (see, for instance, \cite{Felice} and references therein). By using a density profile for the solar interior and its vicinity, it was shown \cite{Hu} that, independently of the particular function  $f(R)$,  the Cassini-mission \cite{Will} constraint on the first PPN parameter ($|\gamma -1|<2.3\times 10^{-5}$) implies that $|f_{R_g}|<4.9\times10^{-11}$. Here, $f_{R_g}=f_R(R=8\pi G\rho_g)$ and $\rho_g\sim 10^{-24} g/cm^3$ is the mean galactic density. Using Eqs. (\ref{DR})
 and (\ref{Rstar})
 it is straightforward to verify that this constraint can easily be satisfied by the $f(R)$ given by Eq. (\ref{fR}). Constraints from violation of the equivalence principle, slightly tighter \cite{Felice}, can also be easily satisfied in $\gamma$ gravity. 
 
 On the other hand, in Ref.~\cite{Hu} there are also arguments indicating that our galaxy halo requires $|f_{R_0}| < 10^{-6}$.  As we will show below, if we are interested in $\gamma$-gravity models whose expansion history can be discriminated from $\Lambda$CDM, this constraint will not be satisfied even for models with $n>1$. However, as also remarked in Ref.~\cite{Hu}, although suggestive, this bound is overrestrictive and should not be considered as definitive. For instance, it depends on when the galactic halo was formed and on the density profiles of the structures in which the galaxy is embedded. As observed in Ref.~\cite{Linder}, the large steepness of the exponential $f(R)$ gravity ($n=1$) ameliorates the situation as compared to many $f(R)$ theories. This occurs because the GR limit of the theory is more easily (i.e, more promptly) recovered when we go from low- to high-curvature regimes. This is more clearly seen if we consider the behavior of $|f_R|$ at structure formation redshifts ($z\sim 1$). Furthermore, since, for $n>1$, $\gamma$ gravity has an even steeper behavior than exponential gravity, we may expect better results in this case. In the following we show that this is indeed the case.

\begin{figure}[t]
\begin{centering}
    \includegraphics[width=0.49\textwidth]{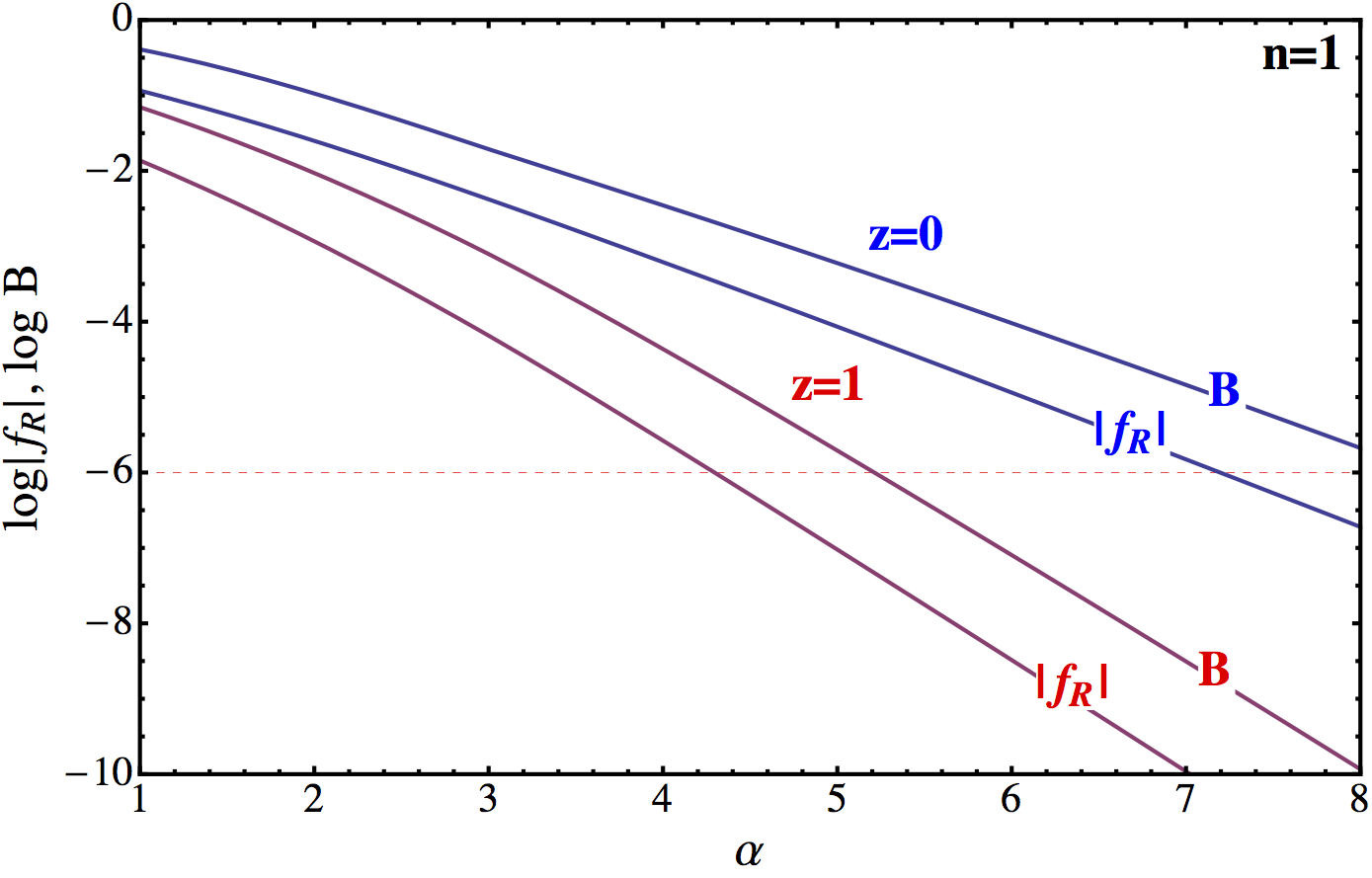}
        \includegraphics[width=0.49\textwidth]{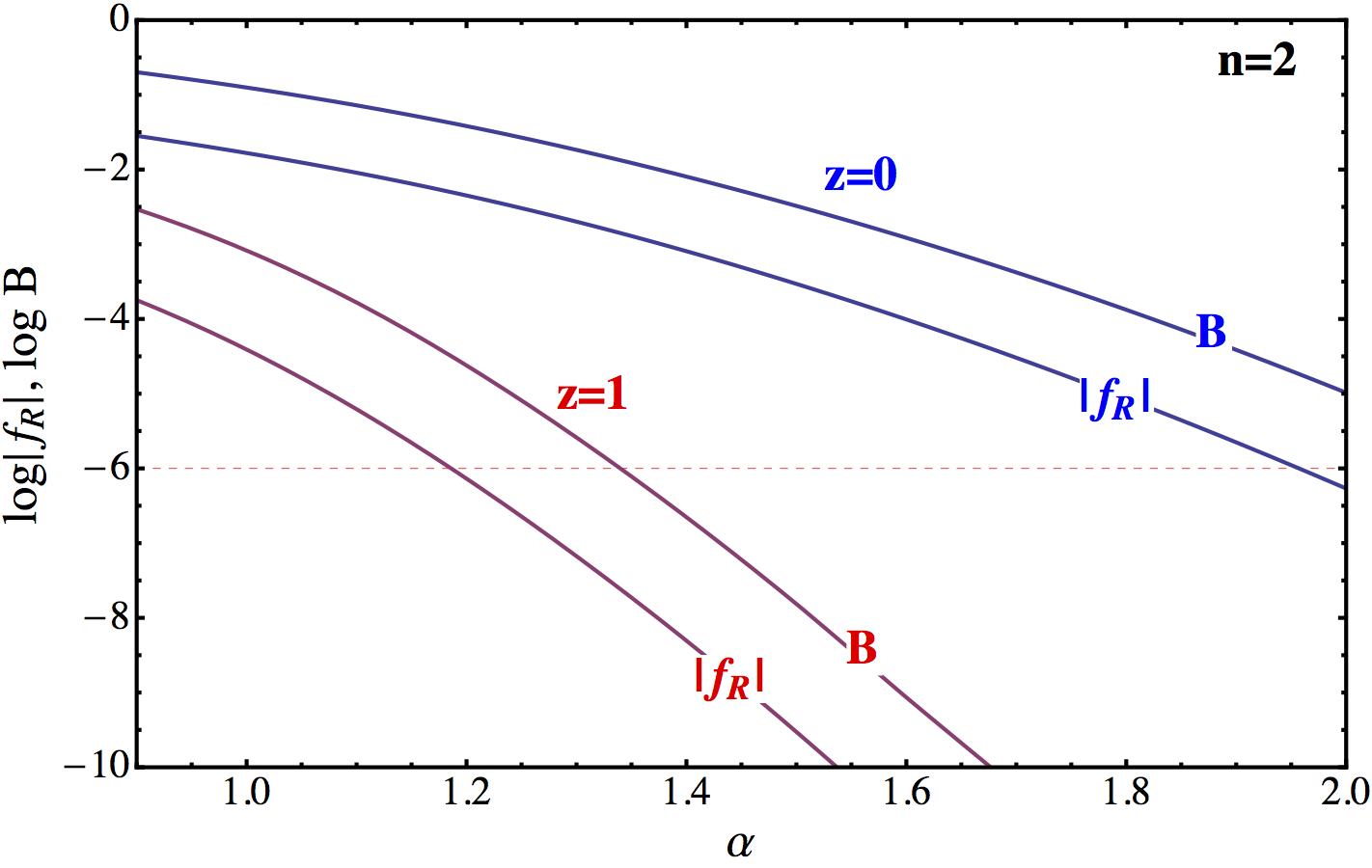}
            \includegraphics[width=0.49\textwidth]{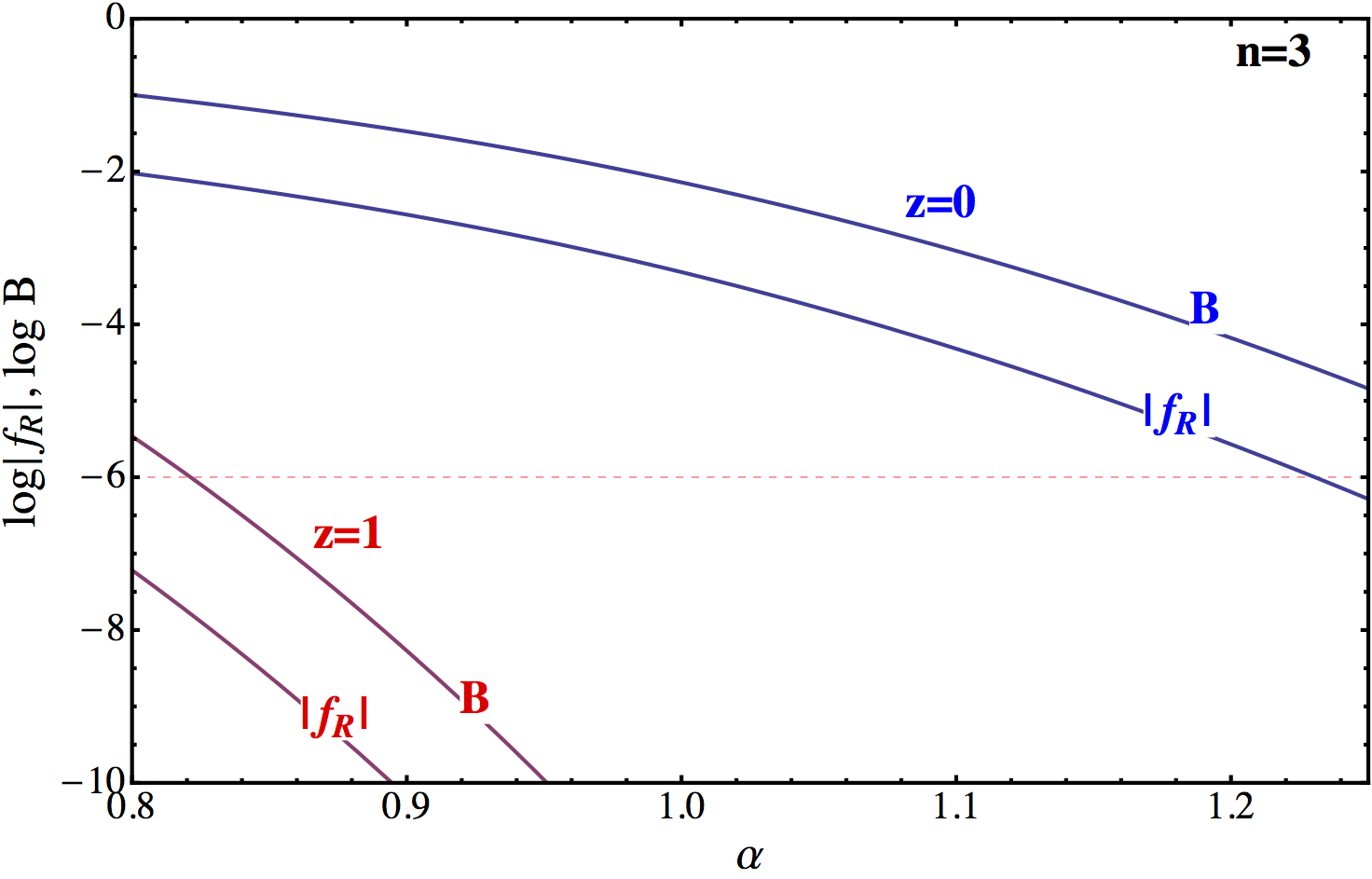}
           \par \end{centering}
    \caption{Behavior of $\log B$ and $\log|f_R|$ with $\alpha$ for $n=1,2$ and $3$. Note the different ranges in the horizontal axes in the different panels. In all of them, the upper (lower) pair of curves indicates the limit at $z=0$ ($z=1)$. The horizontal dashed line indicates the constraint on $f_R$ from the local halo. See text for definitions.}
    \label{fig:7}
\end{figure}

In Fig. \ref{fig:7} we show, for $n=1$, $2$ and $3$, $f_R$ as a function of $\alpha$ at redshifts $z=0$ and $z=1$. We also display in the figure the quantity \cite{Hu}
\begin{equation}
B=\frac{f_{RR}}{1+f_R}R'\frac{H}{H'},
\end{equation}
that is related to the effective Compton wave number ($k_C=aHB^{-1/2}$), above which ($k>k_C$) linear perturbation growth will be affected if $B\gtrsim10^{-5}$. In each panel the upper pair of curves corresponds to $z=0$ while the lower ones correspond to $z=1$.  Observe the strong dependence of $f_R$ and
 $B$ with redshift and how both steepnesses increase with $n$. Since satisfying the condition $\log B < -5 $ is roughly equivalent to  satisfying the constraint $\log|f_R|< -6$, in the following comments we concentrate only on the latter. For $n=1$,  the condition $\log|f_R|< -6$ is satisfied at $z=0$ for $\alpha \gtrsim 7.2$, while $\alpha \gtrsim 4.4$ is required for $z=1$.  The maximum deviation in the effective dark-energy equation of state from a cosmological constant is $|1+w_{de}|_{max}=5 \times 10^{-6}$ at $\alpha = 7.2$, and $10^{-3}$ at $\alpha = 4.4$. So, even if $\alpha = 4.4$, the difference is so small that it will be very difficult to differentiate the expansion history of exponential gravity ($n=1$) from $\Lambda$CDM. For $n=2$, the condition $\log|f_R|< -6$ evaluated at present ($z=0$) requires $\alpha \gtrsim 1.95$ ($|1+w_{de}|_{max}=1.4 \times10^{-5}$), while $\alpha > 1.2$ ($|1+w_{de}|_{max}=1.7 \times10^{-2}$) is required if evaluated at $z=1$.  For $n=3$ the above condition requires $\alpha > 1.24$ ( $|1+w_{de}|_{max}=2.6 \times10^{-5}$) at $z=0$, but is satisfied for any $\alpha > \alpha_{min}$ at $z=1$, having a maximum amplitude  $\simeq 3.7 \times10^{-2}$ at $\alpha = \alpha_{min} \simeq 0.8$. 
Therefore, if the condition $\log|f_R|< -6$ is to be satisfied only at $z=1$, the deviations from $\Lambda$CDM equation of state for $n=2$ models  can be $1$ order of magnitude larger than in the case $n=1$ and could be reached with future surveys \cite{Spergel}. In principle, for larger values of the parameter $n$ the situation should be even better but, as mentioned in the last section, they cannot be much larger than $n=3$ otherwise the transition to the phantom phase will occur too close to the present time and it will be very difficult to detect any deviation in the equation-of-state parameter from $w_{de}=-1$.

\begin{figure}[t]
\begin{centering}
    \includegraphics[width=0.49\textwidth]{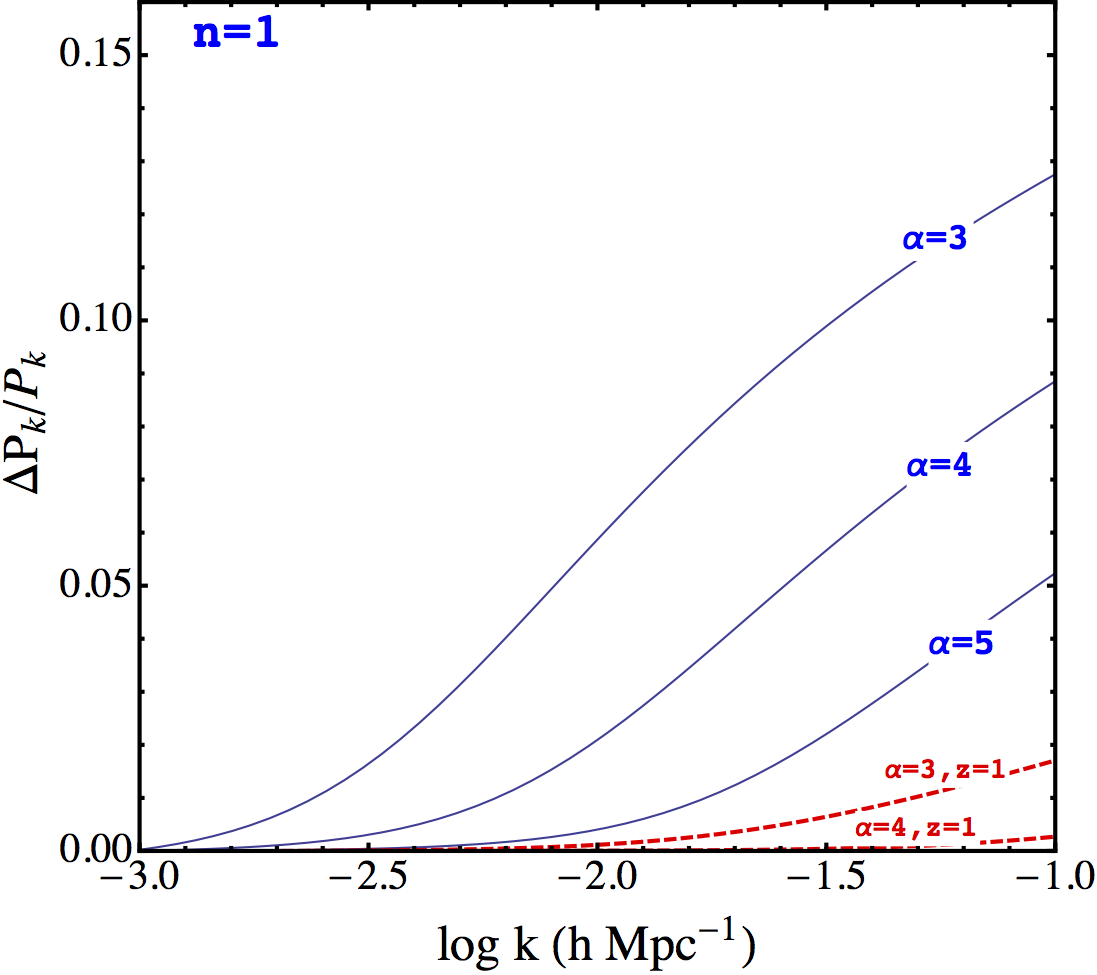}
        \includegraphics[width=0.49\textwidth]{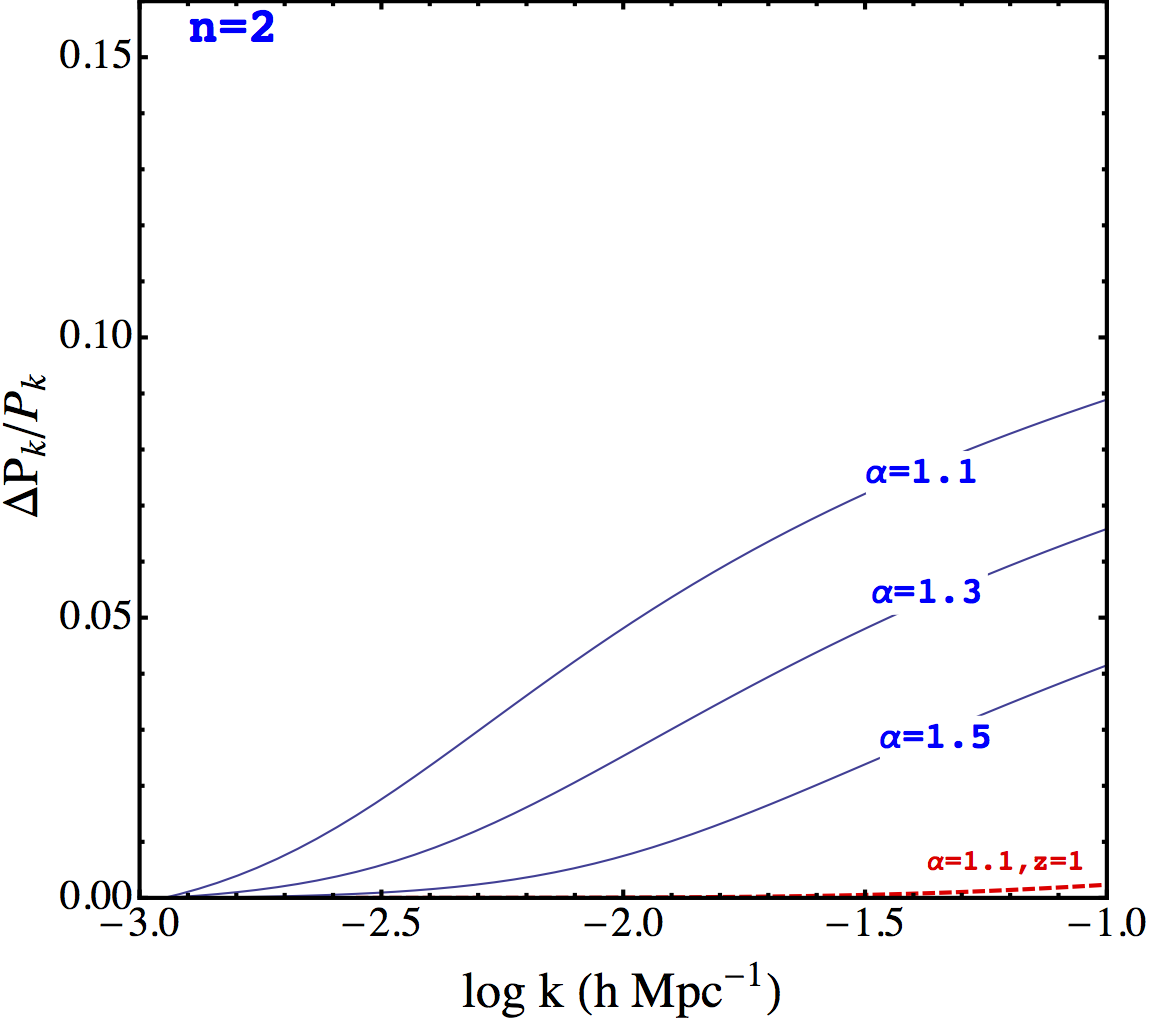}
            \includegraphics[width=0.49\textwidth]{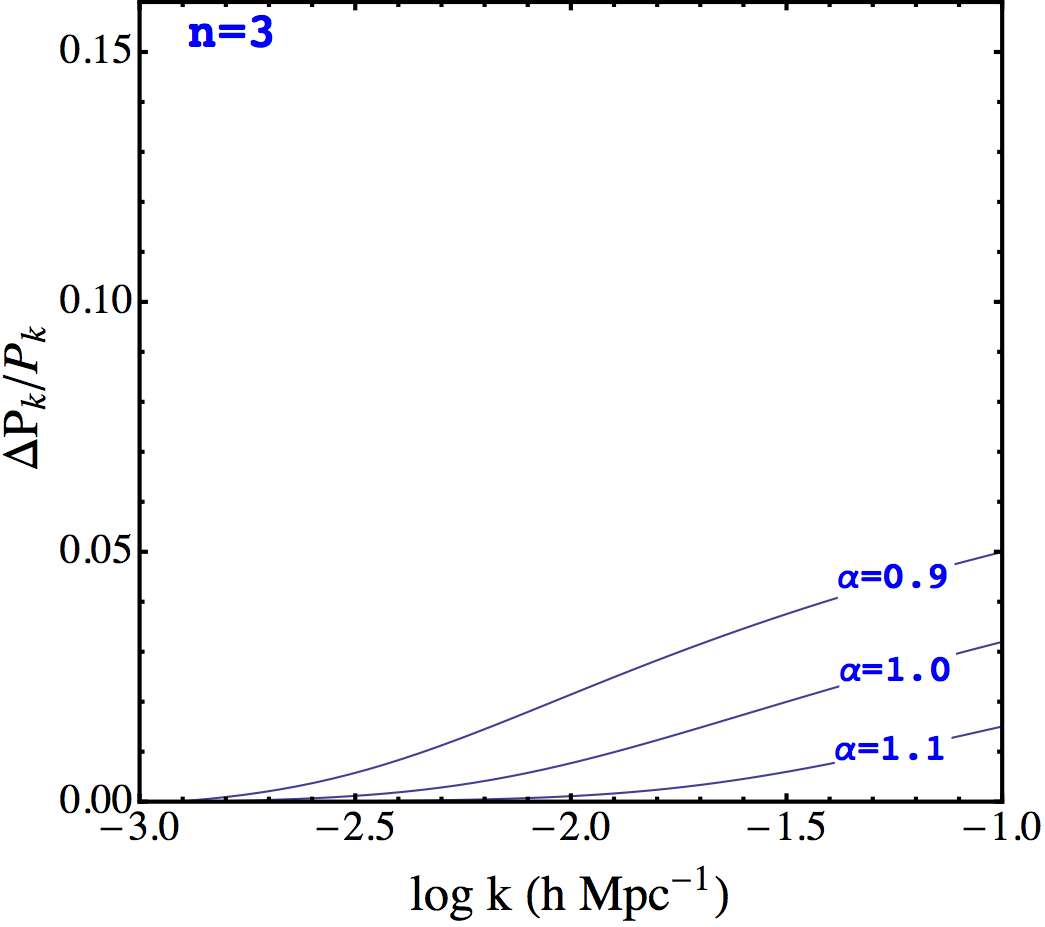}
              \includegraphics[width=0.49\textwidth]{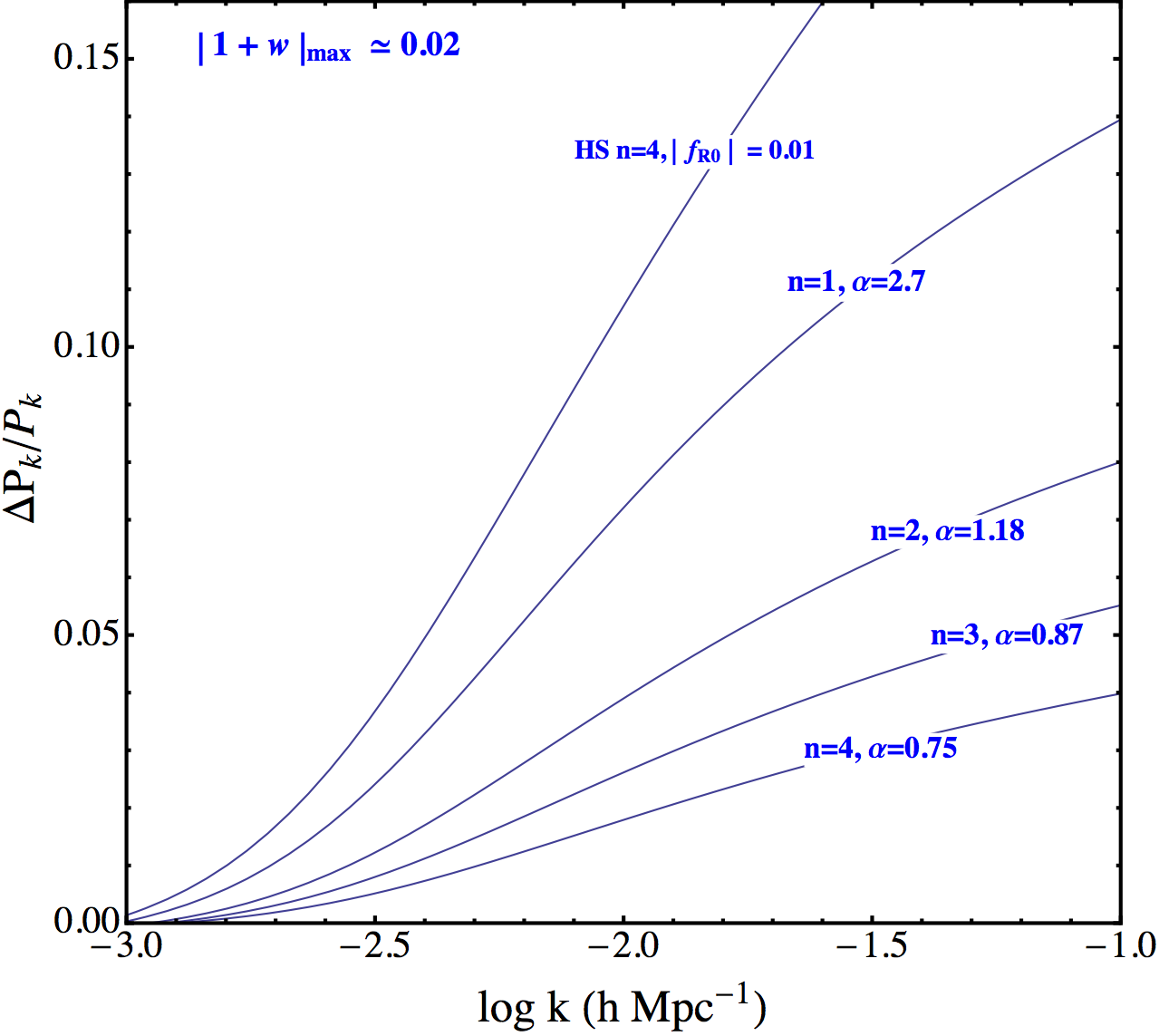}
           \par \end{centering}
    \caption{Fractional change in the matter power spectrum relative to $\Lambda$CDM for different values of $n$ and $\alpha$, as indicated in the plots. Note that for larger $n$ it takes a smaller $\alpha$ to yield a negligible change in the power spectrum. 
    All the curves, except where noted, correspond to $z=0$. In this case we obtain $\log k_{C0}$(hMpc$^{-1}$)$=(-2.62,-2.25,-1.87)$ for $n=1$ and $\alpha=(3,4,5)$;   $\log k_{C0}$(hMpc$^{-1}$)$=(-2.91,-2.61,-2.23)$ for $n=2$ and $\alpha=(1.1,1.3,1.5)$; and  $\log k_{C0}$(hMpc$^{-1}$)$=(-2.74,-2.41,-1.96)$ for $n=3$ and $\alpha=(0.9,1.0,1.1)$. The curves at $z=1$ are only plotted in the first two panels because they coincide with the horizontal axes in the other panels. In the lower right panel, the parameters were chosen so that every curve yields a maximum deviation of $\simeq 2\%$ in the effective equation-of-state parameter $w_{de}$. HS stands for the model introduced in Ref.~\cite{Hu}. }
    \label{fig:8}
\end{figure}

We now consider the linear growth of cosmological matter density perturbations in the subhorizon regime.  In this regime, for $|f_R| \ll 1$, the differential equation for the matter density contrast ($\delta$) can be approximated by \cite{Zhang, Pogosian, Dombriz}
\begin{equation}
\delta''+\delta' \left( 2+\frac{H'}{H} \right) - \delta e^{-3y} \frac{1-2Q}{2-3Q} \frac{3{H_0}^2 \tilde{\Omega}_{m0}}{H^2(1+f_R)}=0
\label{delf}
\end{equation} 
where 
\begin{equation}
Q(k,y)=-\frac{2f_{RR}c^2k^2}{(1+f_R)e^{2y}}.
\end{equation}
In GR, $f_R=Q=0$ and there is no scale dependence for the density contrast in the linear regime. For $w$CDM the growing mode can be expressed in terms of hypergeometric functions $_2F_1$ as  \cite{Silveira}
\begin{equation}
\delta_+ \propto \frac{1}{1+z} \;{_2}F_1\left[ - \frac{1}{3w}, \frac{w-1}{2w}, 1-\frac{5}{6w};-(1+z)^{3w} \frac{1-\tilde{\Omega}_{m0}}{\tilde{\Omega}_{m0}}\right]. \label{delw}
\end{equation}
We solved Eq. (\ref{delf}) numerically and obtained the growing mode for the $f(R)$ given by Eq. (\ref{fR}). By using (\ref{delw}) (with $w=-1$), we then obtained the fractional change in the matter power spectrum $P(k)$ relative to $\Lambda$CDM. 
Figure \ref{fig:8} shows $\Delta P_k/P_k$ at $z=0$ for different values of $n$ and $\alpha$. It is also displayed in the figure (red dashed curves) $\Delta P_k/P_k$ at $z=1$ for $n=1$ ($\alpha=3$ and $4$) and for $n=2$ ($\alpha=1.1$). For higher values of $n$ there is effectively no deviation from $\Lambda$CDM at this redshift. Therefore, it is clear from this figure the effect of the steepness in suppressing deviations from $\Lambda$CDM in the linear mass power spectrum as we go to higher redshifts. For fixed $n$ the suppression also depends on $\alpha$, with smaller deviations occurring for larger $\alpha$. In the lower right panel we display $\Delta P_k/P_k$ at $z=0$ for a fixed value of the maximum deviation in the effective dark-energy equation of state ($|1+w_{de}|_{max}\simeq 2\times 10^{-2}$) for several models. From the figure we see the improvement of exponential gravity ($n=1$) when compared to the Hu and Sawicki (HS) model \cite{Hu} with $\tilde{\Omega}_{m0}=0.28$, $n=4$ and $f_{R_0}=0.01$. 

At this point it is important to make the following remark. In the HS model for fixed $\tilde{\Omega}_{m0}$ and $f_{R_0}$ it is possible to decrease $\Delta P_k/P_k$ if one increases the HS steepness parameter $n$. For instance, for the same $\tilde{\Omega}_{m0}$ and $f_{R_0}$ as above but with $n=10.45$ we obtain $\simeq12\%$ deviation in $\Delta P_k/P_k$ at the smallest linear scale ($k=0.1$ h Mpc$^{-1}$) while keeping  $|1+w_{de}|_{max}\simeq 2\times 10^{-2}$. Furthermore, $|f_R|$ at $z=1$ for this model is comparable to the exponential case with $\alpha \simeq 3$ and, in this sense, there is no significant difference between exponential gravity and the HS model. One can argue that by further increasing the HS steepness parameter it would be possible to obtain even better results as compared to the exponential gravity. However, for fixed $\tilde{\Omega}_{m0}$ and $f_{R_0}$, there is a maximum value of the HS steepness parameter above which the cosmological viability conditions are violated.  For instance, HS models with $\tilde{\Omega}_{m0}=0.28$,  $f_{R_0}=0.01$ and $n\gtrsim10.5$ do not have a late time de Sitter attractor. To reduce $\Delta P_k/P_k$ in the HS model it is necessary to reduce $f_{R_0}$, but in this case one also reduces the maximum deviations in the equation-of-state parameter. What is important to emphasize here is that there is more freedom in $\gamma$ gravity in the sense that is possible to have smaller values for $\Delta P_k/P_k$ while keeping relatively high $w_{de}$ deviations. For instance, as shown in Fig. \ref{fig:8} if $n=3$ and $\alpha=0.87$ the deviation at $k=0.1$ h Mpc$^{-1}$ is less than $6\%$, while for $n=4$ and $\alpha=0.75$ it is $\simeq 4\%$.

\section{Conclusions}\label{conclusions}

It is very difficult to have modified $f(R)$ gravity models that satisfy all the viability and stability criteria, with an effective equation-of-state parameter
distinguishable from $\Lambda$ and being, at the same time, in accordance with both large scale structure formation and local tests of gravity.  In this work, we presented a class of generalized exponential $f(R)$ gravity theory, dubbed $\gamma$ gravity, and investigated  the background expansion history of cosmological models belonging to this class. We examined the behavior of the effective equation-of-state parameter as a function of redshift for different values of the parameters $\alpha$ and $n$, and discussed the constraints from local gravity tests and linear growth of structure. We showed that if the parameter controlling the steepness $n\gtrsim 2$, it is possible to have less than $10\%$ deviations from the $\Lambda$CDM mass power spectrum at all linear scales while having, at the same time, an effective equation of state such that $|1+w_{de}|_{max}\simeq 2-4 \times 10^{-2}$. This does not seem possible in exponential gravity or other viable $f(R)$ theory present in the literature. We argued that  $\gamma$-gravity theory enlarges the spectrum of currently viable $f(R)$ cosmological models opening the possibility of models that could be discriminated from $\Lambda$CDM, not only at the perturbative but also at the background level with foreseeable future experiments. In this sense models with $n \simeq 3$ are the most promising ones.

It should be remarked however, that further investigations are necessary, in particular cosmological simulations in the nonlinear regime should be performed, trying to constrain more the parameter space of  the $\gamma$ gravity theory. Besides, in general the precision with which the equation-of-state parameter will be measured in future dark energy surveys is estimated assuming constant $w_{de}$ or simple redshift-dependent parametrizations. Furthermore, in all our analyses we assumed a constant $\tilde{\Omega}_{m0}$. Therefore, we believe it is important to perform realistic simulations to  quantify the extent in which $\gamma$-gravity theory can give rise to models with a cosmic expansion history observationally distinguishable from $\Lambda$CDM. Investigations in these directions are underway.

\section*{\textbf{Acknowledgements}}

We would like to thank Vin\'{i}cius Miranda for useful discussions and constructive criticisms that helped us to improve this work. M. O. thanks the Brazilian research agency CNPq for support. S.E.J. acknowledges support from ICTP and FAPERJ. I.W. is partially supported by the Brazilian research agency CNPq.

\end{document}